Original Paper

# Enhancing Health Care Accessibility and Equity Through a Geoprocessing Toolbox for Spatial Accessibility Analysis: Development and Case Study

Soheil Hashtarkhani[1], PhD; David L Schwartz[2], MD; Arash Shaban-Nejad[1], MPH, MSc, PhD

[1]Center for Biomedical Informatics, Department of Pediatrics, College of Medicine, The University of Tennessee Health Science Center, Memphis, TN, United States
[2]Department of Radiation Oncology, College of Medicine, University of Tennessee Health Science Center, Memphis, TN, United States

**Corresponding Author:**
Arash Shaban-Nejad, MPH, MSc, PhD
Center for Biomedical Informatics, Department of Pediatrics, College of Medicine
The University of Tennessee Health Science Center
50 N Dunlap Street, R492
Memphis, TN, 38103
United States
Phone: 1 9012875863
Email: ashabann@uthsc.edu

## Abstract

**Background:** Access to health care services is a critical determinant of population health and well-being. Measuring spatial accessibility to health services is essential for understanding health care distribution and addressing potential inequities.

**Objective:** In this study, we developed a geoprocessing toolbox including Python script tools for the ArcGIS Pro environment to measure the spatial accessibility of health services using both classic and enhanced versions of the 2-step floating catchment area method.

**Methods:** Each of our tools incorporated both distance buffers and travel time catchments to calculate accessibility scores based on users' choices. Additionally, we developed a separate tool to create travel time catchments that is compatible with both locally available network data sets and ArcGIS Online data sources. We conducted a case study focusing on the accessibility of hemodialysis services in the state of Tennessee using the 4 versions of the accessibility tools. Notably, the calculation of the target population considered age as a significant nonspatial factor influencing hemodialysis service accessibility. Weighted populations were calculated using end-stage renal disease incidence rates in different age groups.

**Results:** The implemented tools are made accessible through ArcGIS Online for free use by the research community. The case study revealed disparities in the accessibility of hemodialysis services, with urban areas demonstrating higher scores compared to rural and suburban regions.

**Conclusions:** These geoprocessing tools can serve as valuable decision-support resources for health care providers, organizations, and policy makers to improve equitable access to health care services. This comprehensive approach to measuring spatial accessibility can empower health care stakeholders to address health care distribution challenges effectively.

## Introduction

The role of geography in understanding and addressing population health and health inequities is hardly deniable [1-3]. Access to health care is a critical indicator of health care system performance and directly impacts population health and disease burden [4,5]. Improving access to primary care, for instance, has been proven to lead to improved health outcomes and decreased potentially avoidable hospitalizations [6,7]. The concept of access plays a significant role in health services and policy research, including both spatial and nonspatial dimensions. Nonspatial access refers to the factors unrelated to geography that influence access, such as affordability, timeliness, accommodation, acceptability, and awareness [8]. Spatial access, on the other hand, involves the geographic elements that influence the availability and accessibility of health care providers and services [9]. The calculation of spatial accessibility involves considering three key factors: (1) supply, (2) demand, and (3) mobility. Supply relates to the infrastructure's locations (eg, health care providers); demand refers to the locations of individuals who are expected to use the infrastructure (eg, patients); and mobility considers the travel costs between demand and supply locations (eg, driving time) [10]. Identifying areas with limited spatial accessibility enables planners and policy makers to understand the distribution of health service locations and reveal and address spatial inequities [11].

Various methods have been used to evaluate spatial accessibility, including gravity models [12], regional availability models [13], and kernel density models [14]. Among the gravity models, the 2-step floating catchment area (2SFCA) method, initially introduced by Radke and Mu [15] and modified by Luo and Wang [16], has been widely used in the literature for measuring spatial accessibility. The 2SFCA approach measures spatial accessibility through a 2-step procedure based on the interaction between supply and demand within a certain catchment as a ratio of provider-to-population [16]. However, the 2SFCA technique has certain limitations. Locations outside the catchment area are entirely out of access, while population locations within the catchment area are assumed to have equal access to health care providers [17]. To overcome these limitations, Luo and Qi [18] proposed an enhanced version of the 2SFCA method known as the enhanced 2SFCA (E2SFCA) method. This enhanced approach differentiates accessibility within a catchment (usually 3 catchments) by incorporating multiple travel time zones and assigning weights based on a decay function within each catchment [18].

Although the advancements in geographic information system (GIS) technology have made the implementation of the 2SFCA-based models more feasible, researchers with limited GIS expertise still face challenges in gathering, preprocessing, analyzing required data sets, and implementing the model. However, GIS software like ArcGIS provides powerful tools, including model builders and Python programming tools, that enable developers to automate data processing by creating custom geoprocessing toolboxes. In this study, our objective is to develop and share Python script tools for implementing 2SFCA and E2SFCA methods in ArcGIS Pro (Esri). Each toolbox was developed in 2 ways: using a distance buffer and travel time (driving time or walking time) in catchment areas. Additionally, we will present a case study assessing the accessibility of hemodialysis services. In this study, we will measure the accessibility score using the 4 developed tools for the age-adjusted demand population in census tracts of the state of Tennessee to include an important nonspatial factor in the accessibility score.

## Methods

### Theoretical Framework and Conceptual Explanations

A total of 4 geoprocessing tools were developed to work in ArcGIS Pro software based on 2SFCA [16] and E2SFCA [18] approaches with the following names:

1. 2SFCA01: 2SFCA with buffer distance catchments
2. E2SFCA01: E2SFCA with buffer distance catchments
3. 2SFCA02: 2SFCA with travel time catchments
4. E2SFCA02: E2SFCA with travel time catchments

After presenting the theoretical background of the 2SFCA and E2SFCA approaches, the development framework for each tool will be presented.

### Models' Theory

#### Theoretical Background of 2SFCA

The 2SFCA method assesses the relationship between resource availability and demand population distribution in 2 steps, resulting in an access score for each demand area.

Step 1: For each facility location (j), identify all demand locations (k) that fall within a specified catchment area ($d_0$) from the facility. The provider-to-demand ratio ($R_j$) within the catchment area is calculated using equation 1:

$$R_j = \frac{S_j}{\sum_{k \in \{d_{kj} \leq d_0\}} P_k} \quad (1)$$

where $P_k$ represents the population at demand location k within catchment area j ($d_{kj} \leq d_0$), $S_j$ is the capacity or number of providers at location j, and $d_{kj}$ is the distance (or travel time) between k and j.

Step 2: For each demand location i, search for all facility locations (j) within the specified catchment area ($d_0$) from location i, and calculate the summed provider-to-demand ratios ($R_j$) obtained in step 1 using equation 2:

$$A_i^F = \sum_{j \in \{d_{ij} \leq d_0\}} R_j \quad (2)$$

In equation 2, $A_i^F$ represents the accessibility at demand location i based on the 2SFCA method. $R_j$ denotes the provider-to-demand ratio at facility location *j* that falls within the catchment area of the demand location *i* (ie, $d_{ij} \leq d_0$), and $d_{ij}$ is the distance (or travel time) between *i* and *j*.

#### Theoretical Background of E2SFCA

The classic 2SFCA relies on a dichotomous distance decay function, assuming that individuals within catchment areas have equal access to services, while those outside catchment areas have no access at all. To overcome the distance decay limitation of the classic 2SFCA, we also used the E2SFCA procedure introduced by Luo and Qi [18] as follows:

In step 1, for each facility location (j), 3 distance or travel time catchment areas are created, including zone 1: 0-5 miles (0-8 km); zone 2: 5-10 miles (8-16 km); and zone 3: 10-15 miles (16-24 km) or minutes. Search all demand locations (k) that were within the zones ($D_r$) for location j and compute the weighted provider-to-demand ratio ($R_j$) using equation 3:

$$R_j = \frac{S_j}{\sum_{k \in \{d_{kj} \in D_r\}} P_k W_r} = \frac{S_j}{\sum_{k \in \{d_{kj} \in D_1\}} P_k W_1 + \sum_{k \in \{d_{kj} \in D_2\}} P_k W_2 + \sum_{k \in \{d_{kj} \in D_3\}} P_k W_3} \quad (3)$$

where $P_k$ is part of the demand k falling within the catchment j ($d_{kj}$ $D_r$), $S_j$ is the capacity or number of providers at facility j, $d_{kj}$ the distance (or travel time) between k and j, and $D_r$ is the $r^{th}$ catchment zone (r {1,2,3}) within the catchment. $W_r$ is the distance weight for the $r^{th}$ zone calculated from the Gaussian function capturing the distance decay of access to the facility j.

Step 2: For each demand location i, search all facility locations (j) within the distance (or travel time) threshold of the location i, and summed up the provider-to-population ratios $R_j$ (calculated in step 1) as follows:

$$A_i^F = \sum_{j \in \{d_{ij} \leq D_r\}} R_j W_r = \sum_{j \in \{d_{ij} \in D_1\}} R_j W_1 + \sum_{j \in \{d_{ij} \in D_2\}} R_j W_2 + \sum_{j \in \{d_{ij} \in D_3\}} R_j W_3 \quad (4)$$

where $A_i^F$ represents the accessibility at the demand location *i*, $R_j$ is the provider-to-demand ratio at facility location *j* that falls within the catchment of demand location *i* (ie, $d_{kj}$ $D_r$), and $d_{ij}$ *is* the distance (or travel time) between *i* and *j*. The same distance weights derived from the Gaussian function used in step 1 are applied to each zone to account for the distance decay.

## Tool Frameworks

### Overview

The accessibility tools in this study use 2 spatial data sets as input, provider and population data, and 1 output data set, as described in the following paragraphs. The framework for developing each tool is described in their respective subsections.

### Provider Data

These are point data showing the location of health service providers (eg, hospitals) and contain an ID field and a capacity field. The capacity field is a numeric field including the number of providers (eg, number of physicians) or the number of resources (eg, number of hospital beds).

### Population Data

These are polygon data that contain the geographical areas where the accessibility score is supposed to be calculated (eg, census tracts) and contain an ID and population field. The population field is a numerical field that represents the demand for service. It could be the total population of a census tract or the number of women, children, or older adults.

### Output Feature Class

This polygon feature class is exactly similar to the population input feature class, with 1 added field named "final index." This field demonstrates the accessibility score of service providers in the input regions.

### Framework for Developing 2SFCA01

Figure 1A shows the simplified procedure used to develop the 2SFCA01 tool. Some preprocessing steps, including duplicating the input data file and creating temporary fields, have been done for each input. Then, in step 1, using the input parameters from the user, the output is created, named "Step 1 output." This feature class is similar to input provider data, with a new field named provider to demand representing the ratio of providers to the demand population in the catchment of each provider facility. Step 2 is relatively straightforward and uses the same buffer size as step 1 to sum up the provider-to-demand values calculated in step 1 for each population area.

Figure 2A shows the screenshot of the 2SFCA01 tool. Users can easily select the input data and fields using combo boxes. As the value of the accessibility score is usually very small, the "per capita" parameter multiplies the score with a user-defined value.

**Figure 1.** Flowchart diagram for creating (A) 2-step floating catchment area (2SFCA) and (B) enhanced 2-step floating catchment area (E2SFCA01) spatial accessibility tools.

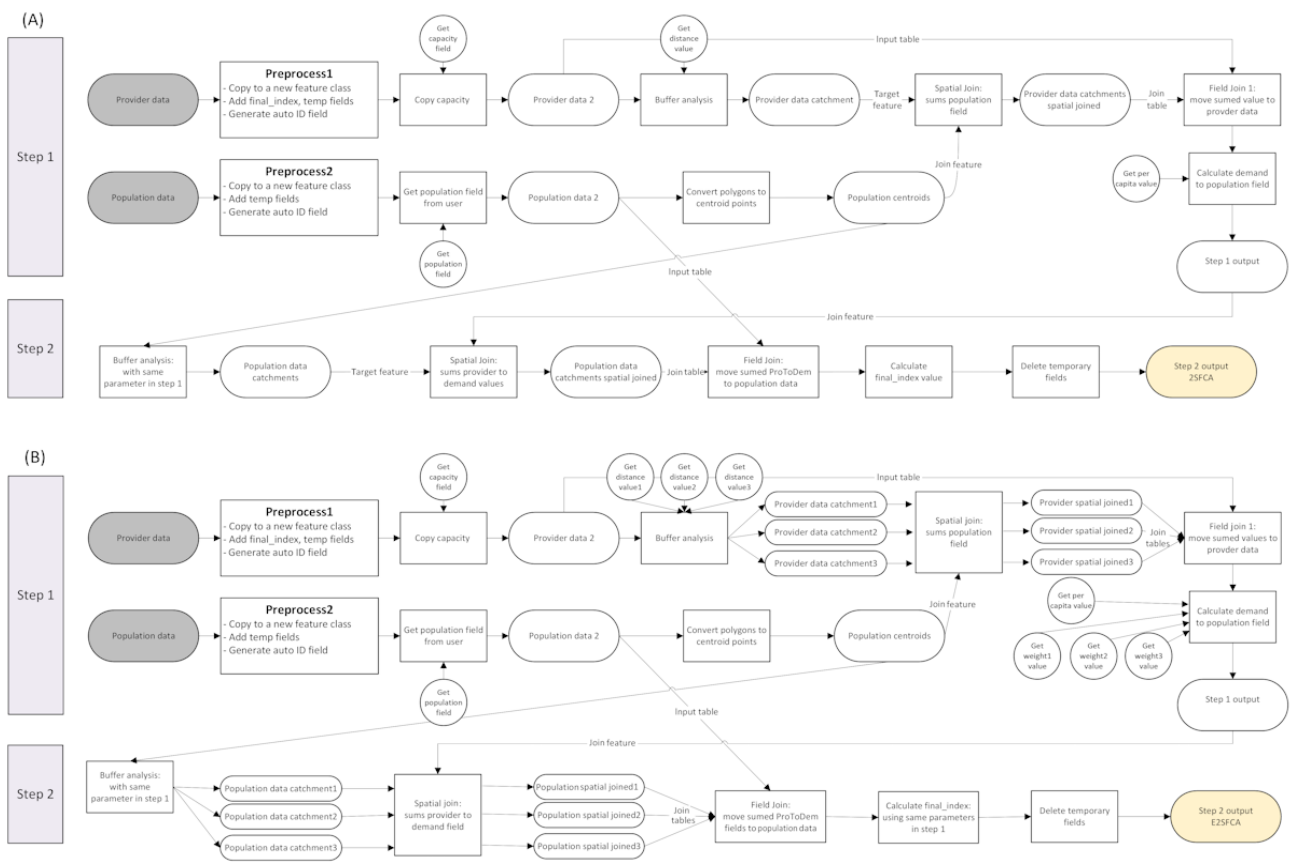

**Figure 2.** Screenshots of the accessibility tools with distance buffer catchments: (A) 2-step floating catchment areas with buffer distance catchments (2SFCA01) and (B) enhanced 2-step floating catchment areas with buffer distance catchments (E2SFCA01).

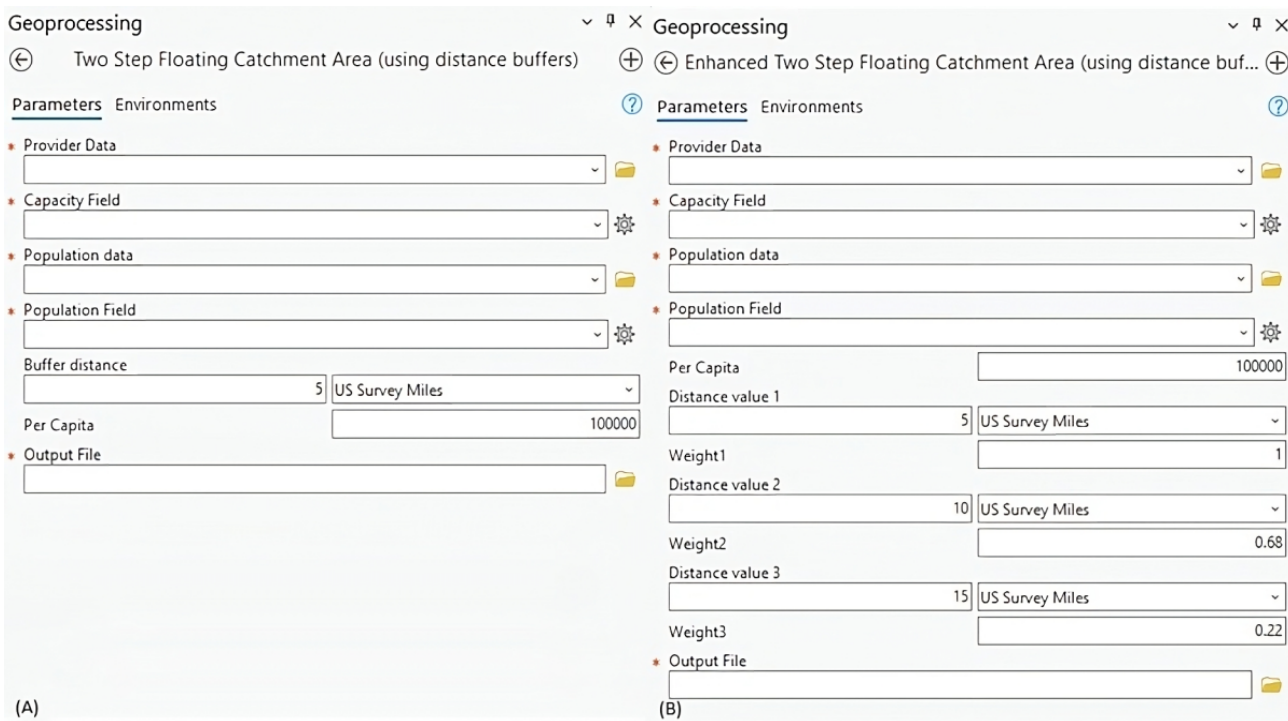

## Framework for Developing E2SFCA01

The procedure to create the E2SFCA01 tool is similar to the 2SFCA01 tool (Figure 1B). As in this model, 3 catchments are necessary for each step, and more parameters from users are required to be defined (circle shapes). The calculated values for each catchment size are combined using user-defined weights. In this tool, the output file not only includes the final index value but also includes the accessibility values for each of the 3 catchment sizes.

Figure 2B shows a screenshot of the toolbox. The default values for distance values are 5, 10, and 15 miles (8, 16, and 24 km) and weights: 1, 0.68, and 0.22, respectively. These default weight sets for distance decay were derived from the original study that developed the model [18], but users should choose the proper weights regarding the purpose and context of the study.

(a)

### Framework for Developing 2SFCA02

The accessibility tools developed with travel time catchments need 2 separate tools to conduct the analysis. The first user should create travel time catchment data for input data using a tool that we developed named "create travel time catchment areas" and then calculate the accessibility index using the 2SFCA02 tool. In order to run the "create travel time catchment areas" tool, users have 2 options. First, if they have a network data set for their study area, they can use it in this tool to create the necessary catchments. If not, they can select ArcGIS Online resources to capture the driving or walking time catchment areas. In this way, the user should have a Network Analysis license with enough credits to use this tool. As shown in Figure 3A, in addition to provider and population data, the ID fields should be defined by the user in this tool. It is necessary to include the output of this tool in the 2SFCA02 and E2SFCA02 tools as input. A value of 10 minutes of driving time is used as the default value, but users can change it to walking distance. Also, the time, date, and direction of the travel can be customized to consider the traffic, as the catchment sizes will be smaller during rush hour than at other times of the day. Two output files, 1 for the provider and 1 for the population data, will be exported. The main 2SFCA02 tool is shown in Figure 3B. This tool has 4 input files: 2 for the provider and population data, and 2 for the travel time catchments derived from the previous step. The ID fields that are specified in the "create travel time catchment area" tool should be specified. Other features of the toolbox and the process are similar to 2SFCA01.

**Figure 3.** Screenshots of the proposed tools: (A) Create Travel Time Catchments tool, (B) the 2-step floating catchment areas with travel time catchments (2SFCA02) tool, and (C) the enhanced 2-step floating catchment areas with travel time catchments (E2SFCA02) tool.

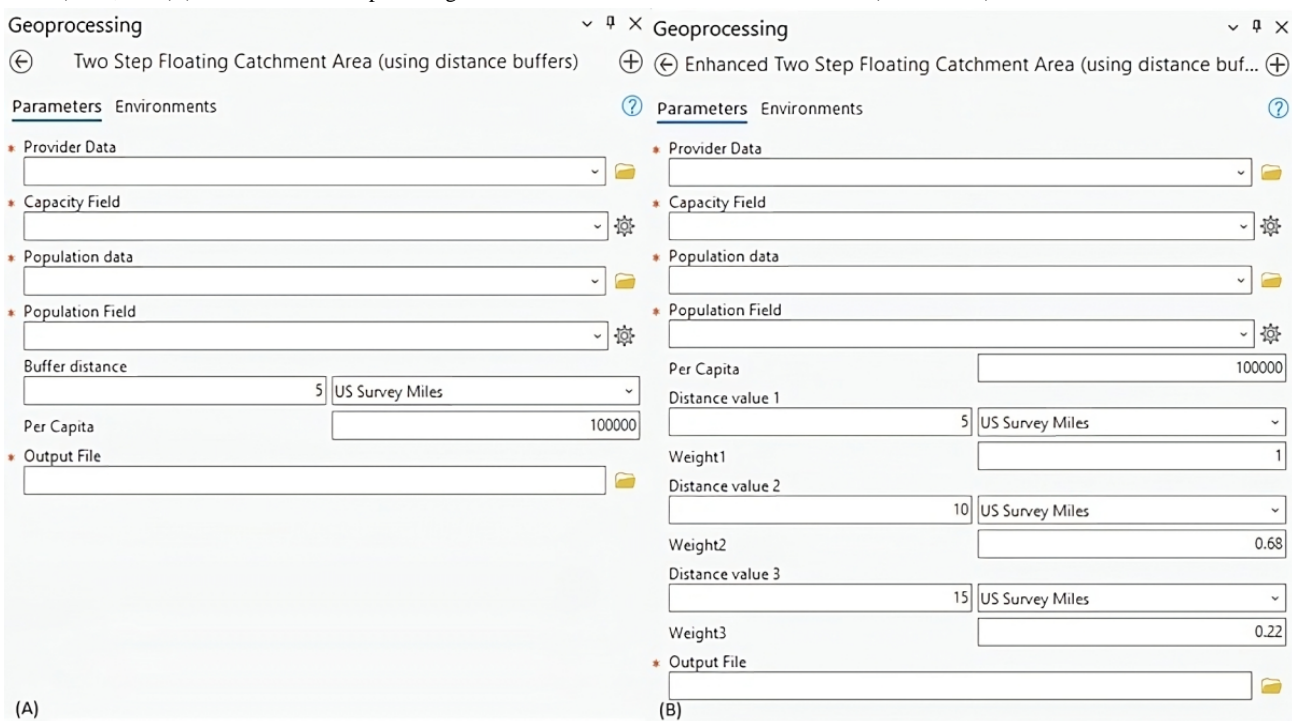

### Framework for Developing E2SFCA02

To use this tool, the user should use the create travel time catchment areas tool similar to adding 3 travel time values, for example, 10 minutes, 15 minutes, and 20 minutes of driving time. This will create 3 travel time rings around each provider facility and demand population point. Figure 3C demonstrates the E2SFCA02 toolbox to create accessibility.

Python scripts for all of the tools are available in Multimedia Appendix 1, and our implemented executable tools are available for download elsewhere [19].

### Ethical Considerations

This study and the development of the presented tool did not involve data that requires ethical oversight. The case study in this manuscript used publicly available data focusing on dialysis center locations and population age distribution at the census tract level. This secondary analysis of aggregated, open-source data is exempt from institutional review board review, in accordance with the Federal Policy for the Protection of Human Subjects (45 CFR 46).

## Results

**Case Study: Access to Hemodialysis Services in the State of Tennessee**

To demonstrate the practicality of the proposed tools in real-world scenarios, we aimed to assess the accessibility of hemodialysis services in different areas of the state of Tennessee.

Hemodialysis is a crucial treatment for individuals with end-stage kidney disease (ESKD), as it eliminates waste products and extra fluid from the blood when the kidneys can no longer do this on their own. Without this treatment, people with ESKD would quickly develop life-threatening complications. This is why the geographical accessibility of hemodialysis services is a critical issue. A study showed that patients living 60 minutes away from a hemodialysis center not only run an increased risk of mortality but also have a significantly lower quality of life compared with patients living 15 minutes or less away [20].

**Input Data Sets**

*Provider Data*

The address and location of hemodialysis centers and the number of machines in each center have been extracted from the Centers for Medicare and Medicaid Services [21]. The address data for the 192 hemodialysis centers have been geocoded into coordinates using the ArcGIS world geocoding service. The resulting shapefile includes ID and capacity (number of machines) fields imported to ArcGIS Pro software.

*Population Data*

The shapefile of census tracts in the state of Tennessee and their total population has been downloaded from the US Census Bureau website [22]. The state of Tennessee includes 1701 census tracts with an average population of 3981 (SD 1646) people. To have a better proxy of target demand, we adjusted the total population with the age distribution of each census tract. Researchers can adjust the target population based on demand and health care needs [23]. We derived the incidence rates of ESKD for various age groups from the 2020 report by the US Department of Health and Human Services [24]. According to the report, the incidence rate of ESKD among individuals aged between 0 and 12 years is 11 cases per million, whereas it is 2080 cases per million for those aged 75 years or older. Table 1 details the age-adjusted demand for ESKD health care services for each census tract, calculated using the following formula: Age-adjusted demand = $(N_{0-17} \times 1) + (N_{18-44} \times 7) + (N_{45-64} \times 51) + (N_{65-74} \times 106) + (N_{\geq 75} \times 189)$ where $N_{a-b}$ represents the number of individuals in the census tract aged a to b years.

The shapefile of census tracts in Tennessee, including GeoID as an ID field and age-adjusted demand as a population field, was imported into ArcGIS Pro software. Figure 4 demonstrates the location of hemodialysis centers and the population density of census tracts in Tennessee.

**Table 1.** End-stage kidney disease (ESKD) incidence rate in different age groups of the US population and calculated weighted demand values.

| Age group (years) | ESKD incidence rate per million, n | Weight |
| --- | --- | --- |
| 0-12 | 11 | 1 (baseline) |
| 18-44 | 77 | 7 |
| 45-64 | 561 | 51 |
| 65-74 | 1171 | 106 |
| ≥75 | 2080 | 189 |

**Figure 4.** The locations and distribution of hemodialysis centers in Tennessee; 1 mile=1.6 km.

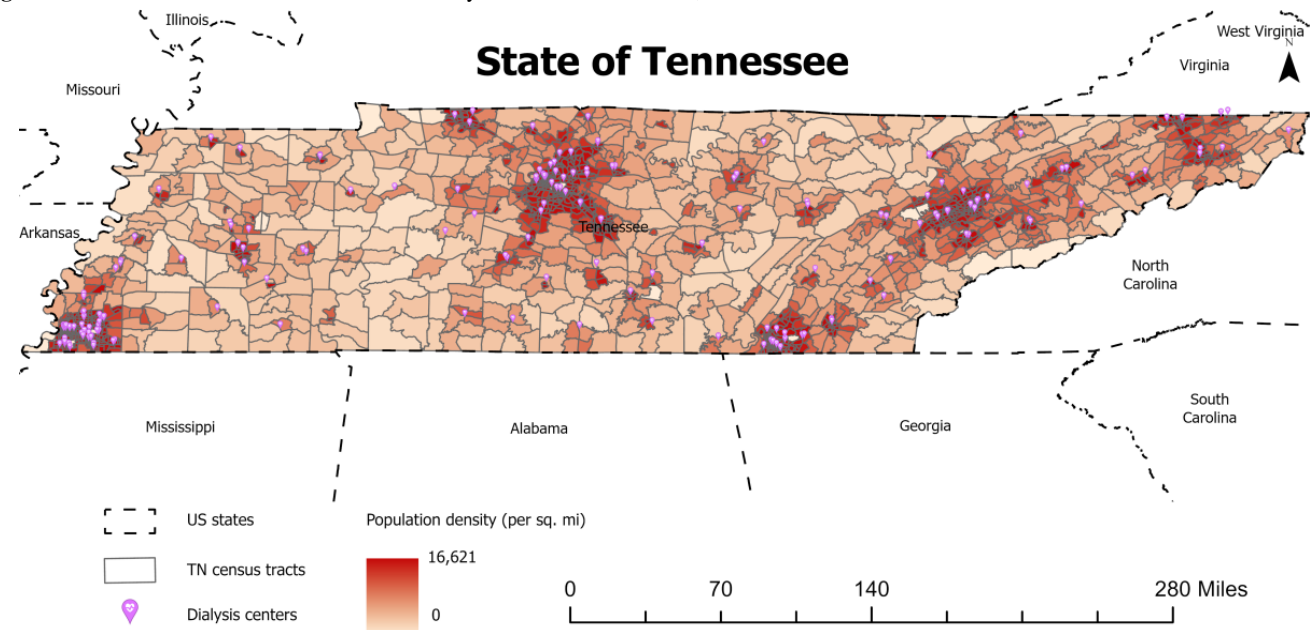

## Mobility

To measure the accessibility indexes, we used a distance buffer size of 15 miles (24 km) for the 2SFCA01 tool and 5, 10, and 15 miles (8, 16, and 24 km) for E2SFCA01. Also, for the 2SFCA02 tool, we used 30 minutes' drive-time catchments and 10, 20, and 30 minutes for the E2SFCA tool. The distance decay sets of 1, 0.68, and 0.22 have been used for weighting each catchment in enhanced versions. The resulting accessibility index for each tool was symbolized in a geographical map using natural break classification (Figure 5).

As depicted in Figure 5, each tool generates distinct accessibility scores in the state of Tennessee, although the overall trends remain largely consistent. Rural and suburban areas generally exhibit lower access scores compared to urban areas, where a concentration of hemodialysis centers is observed. The 2SFCA tool's findings reveal that in areas with high access for every 100,000 people, there are between 12.9 and 27.7 dialysis machines available within a 15-mile (24-km) radius or from 15.7 to 21.2 machines accessible within a 30-minute travel time. Conversely, in regions with low access, the availability of these resources is nearly nonexistent. Interpreting the results from the E2SFCA tool is not straightforward due to its weighted measurement approach. Notably, it is evident that areas represented by white, indicating a lack of health care resources within the distance thresholds, should be prioritized for informed resource allocation efforts [25].

**Figure 5.** Access to hemodialysis centers in Tennessee using the following tools: (A) 2-step floating catchment areas with buffer distance catchments (2SFCA01), (B) enhanced 2-step floating catchment areas with buffer distance catchments (E2SFCA01), (C) 2-step floating catchment areas with travel time catchments (2SFCA02), and (D) enhanced 2-step floating catchment areas with travel time catchments (E2SFCA2); 1 mile=1.6 km.

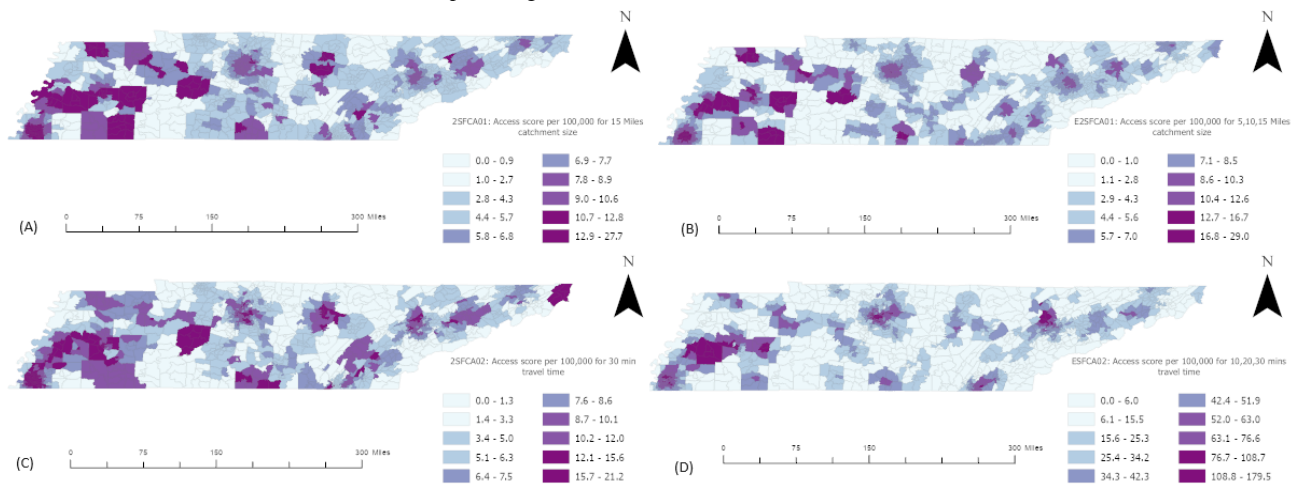

## *Discussion*

### Overview

The primary objective of this study was to introduce GIS tools for measuring the accessibility of health care resources, specifically focusing on the widely used 2SFCA model and its enhanced versions. 2SFCA is the most popular model for measuring the accessibility of health care resources in the literature, and many extensions have been introduced to improve its functionality [26,27]. Our proposed tools aim to assist the health research community in identifying underserved areas in terms of health care accessibility. The development of these tools can significantly streamline the process of assessing and addressing spatial disparities in health care access.

The classic 2SFCA tools (2SFCA01 and 2SFCA02) offer the advantage of simplicity in interpretation. For policy makers, an access score of 20 per 100,000 with a 60-minute catchment size means that there are 20 health care providers accessible for every 100,000 individuals within a 60-minute drive time. This straightforward interpretation facilitates policy makers' understanding of accessibility. On the other hand, the enhanced versions (E2SFCA01 and E2SFCA02) use weighted scores in each step, making the interpretation more complex. However, the use of a distance decay function in the enhanced versions helps overcome the limitations of the classic 2SFCA model and makes it more accurate for comparing the accessibility of different regions.

The use of travel time catchments in 2SFCA02 and E2SFCA02 tools has several advantages. First, travel time offers a more precise measure of accessibility as it takes into account factors such as traffic congestion, road type, and urbanization factors. It can also accommodate different modes of mobility, including walking time. However, it is essential to note that using travel time tools requires access to a Network Analysis license with sufficient ArcGIS Online credits. Each travel time calculation consumes approximately 0.5 ArcGIS Online credits. In the case study, we analyzed the E2SFCA02 tool with 2800 credits, considering 1701 census tracts and 192 hemodialysis centers in Tennessee.

We introduced the Create Travel Time Catchments tool as a standalone prerequisite for 2SFCA02 and E2SFCA02. This approach enables users to generate catchment areas for health care providers and the population, facilitating multiple runs of the access model without incurring additional time and cost for the initial step. To make travel time calculations more adaptable, we designed the tool with flexible options. Users with a network data set covering their study area can compute travel time catchments without consuming ArcGIS Online credits. For those lacking a local network data set but having sufficient ArcGIS Online credits, the tool can leverage web-based resources. However, in the absence of both a network data set and ArcGIS Online credits, the 2SFCA01 and E2SFCA01 tools are viable alternatives, as they use simple Euclidean distance buffers for analysis. Ideally, the E2SFCA02 tool, which incorporates both distance decay and travel time catchments, offers a more realistic measure that closely mirrors real-world health care accessibility dynamics.

We could not identify any peer-reviewed studies presenting a comprehensive spatial accessibility toolbox in ArcGIS. However, there have been a few attempts documented in the gray literature. One such effort was made by Langford et al [28], who shared a tool named USW-FCA2 on ResearchGate using an E2SFCA model. Their tool requires a network data set for the study area and a Network Analysis license in ArcMap. In comparison, our toolbox offers significantly more functionalities and options that cater to the specific needs of the target users. Additionally, some studies have used spatial accessibility tools on alternative platforms. Saxon et al [29] developed an open software environment based on the Python-based *PySaL* package for measuring spatial accessibility. They calculated travel costs by incorporating precomputed origin-destination distance matrices for all US census tracts and census blocks in the 20 major cities.

In the case study, we demonstrated the integration of a nonspatial factor, age, with spatial accessibility. Age is an important determinant of health care demand, and regions with older populations tend to have a higher demand for health services [10]. The procedure used in the case study to adjust the age of the demand population can be extended to consider other factors such as ethnic groups or disease distributions. The resulting geographical maps revealed disparities in access to hemodialysis services across the state of Tennessee. Urban areas, where hemodialysis centers are concentrated, generally exhibited higher accessibility scores. However, areas in the southern parts of the state displayed lower accessibility scores, indicating a need for attention and prioritization in resource allocation.

This study does have limitations to consider. The developed tools are designed specifically for ArcGIS Pro, which may limit their usability for researchers using other software platforms like QGIS. However, future studies could explore the adaptation of these tools to different GIS software to ensure broader accessibility and usability for researchers across various platforms.

### Conclusion

In conclusion, the developed tools for measuring the accessibility of health care resources offer valuable benefits to researchers across various domains. For large-scale analyses, such as country-level assessments, the 2SFCA01 and E2SFCA01 tools provide fast analysis with basic software requirements, making them accessible and efficient options. Additionally, the 2SFCA02 and E2SFCA02 tools offer a more realistic measure of accessibility by incorporating travel time catchments that consider traffic and transportation modes. Among these tools, the E2SFCA02 tool stands out as a powerful option as it considers both distance decay and uses travel time catchments, providing a comprehensive approach to measuring health care accessibility. Overall, these tools empower policy makers and researchers to gain valuable insights into identifying underserved areas and formulating effective resource allocation strategies. By assessing spatial disparities in health care access, these tools contribute to improving equity and enhancing health care service delivery. In the future, we will use and evaluate the outputs of this study for various health resource allocation projects, including primary care providers and cancer care services (eg, radiotherapy). Furthermore, we will explore the impact of nonspatial factors such as ethnicity, income levels, and different social determinants of health to better understand their contributions to health care accessibility.

### Acknowledgments

This study did not receive any specific grants from funding agencies in the public, commercial, or not-for-profit sectors.

### Data Availability

The data sets generated during and/or analyzed during this study are available through Arcgis' website [19].

### Multimedia Appendix 1

Python scripts for the tools used in the study.
[ZIP File (Zip Archive), 11 KB-Multimedia Appendix 1]

## Abbreviations

**2SFCA:** 2-step floating catchment area
**2SFCA01:** 2-step floating catchment areas with buffer distance catchments
**2SFCA02:** 2-step floating catchment areas with travel time catchments
**E2SFCA:** enhanced 2-step floating catchment area
**E2SFCA01:** enhanced 2-step floating catchment areas with buffer distance catchments
**E2SFCA02:** enhanced 2-step floating catchment areas with travel time catchments
**ESKD:** end-stage kidney disease
**GIS:** geographic information system